\documentclass[11pt]{article}

\usepackage{jheppub_mod}

\usepackage[english]{babel}
\usepackage{amsmath}
\usepackage{amssymb}
\usepackage{amsthm,amsfonts}
\usepackage{dsfont}
\usepackage{tensor}
\usepackage[dvipsnames]{xcolor}
\usepackage[mathstyleoff]{breqn}
\usepackage{slashed}
\usepackage{stmaryrd}
\usepackage{enumerate}
\usepackage[normalem]{ulem}
\usepackage{tikz}
\usepackage[mathscr]{euscript}
\usepackage{mathrsfs}
\usepackage{hyperref}
\usepackage{amsthm}
\usepackage{svrsymbols}
\usepackage{phaistos}
\usepackage{mathtools}
\usepackage{stmaryrd} 
\usepackage{graphicx}

\usepackage{wasysym}
 \DeclareMathAlphabet{\mathpzc}{OT1}{pzc}{m}{it}
 
\usepackage{siunitx}
\usepackage{array}
\usepackage{multirow}
\usepackage{gensymb}
\usepackage{tabularx}
\usepackage{extarrows}
\usepackage{booktabs}
\usetikzlibrary{fadings}
\usetikzlibrary{patterns}
\usetikzlibrary{shadows.blur}
\usetikzlibrary{shapes}

\title{Integrable defects and B\"acklund transformations in Yang-Baxter models}
\author{Saskia Demulder$\,{}^a$,}
\emailAdd{sademuld@mpp.mpg.de}

\author{Thomas Raml$\,{}^{a,b}$}
\emailAdd{thomas.raml@physik.uni-muenchen.de}
\affiliation[a]{Max-Planck-Institut f\"ur Physik, F\"ohringer Ring 6, 80805 M\"unchen, Germany.}
\affiliation[b]{Ludwig-Maximilian-Universit{\"a}t, Theresienstra\ss e 37, 80333 M{\"u}nchen, Germany.}

\abstract{We explore two distinct methods to introduce integrable defects in a family of integrable sigma-models known as Yang-Baxter models. The first method invokes a modified monodromy matrix encoding an integrable defect  separating two integrable systems. As an example we construct integrable defects in the ultralocal version of the $S^2$ Yang-Baxter model or 2d Fateev sausage model. The second method is based on the so-called ``frozen'' B\"acklund transformations. Lifting the construction to the Drinfel'd double, we show how defect matrices can be constructed for inhomogeneous Yang-Baxter models. We provide explicit expressions for the $SU(2)$ non-split Yang-Baxter model for this class of integrable defects. 
}

\begin{document}
	\maketitle

\section{Introduction}
Defects are internal boundaries separating different regions of a bulk theory. When crossing the defect, the fields of the system undergo a discontinuous jump but preserve a particular set of properties. Of particular interest are defects preserving properties that enable us to keep computational control over the total system. In the context of integrable models, that property is integrability.  Adding defects to integrable models is especially interesting considering that they can help us move towards less symmetric or `ideal' configurations and consequently allows one to probe new features in these systems.  Here, we will address the question whether it is possible at all to introduce defects in a class of integrable classical  field theories that are deformed and thus already feature a reduced set of (global) symmetries.

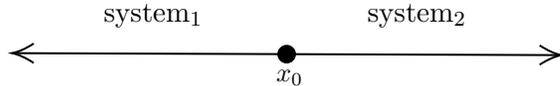
\begin{figure}
\centering
\tikzset{every picture/.style={line width=0.75pt}} 

\begin{tikzpicture}[x=0.75pt,y=0.75pt,yscale=-1,xscale=1]

\draw    (225.24,129.3) -- (498.24,130.28) ;
\draw [shift={(500.24,130.29)}, rotate = 180.21] [color={rgb, 255:red, 0; green, 0; blue, 0 }  ][line width=0.75]    (10.93,-4.9) .. controls (6.95,-2.3) and (3.31,-0.67) .. (0,0) .. controls (3.31,0.67) and (6.95,2.3) .. (10.93,4.9)   ;
\draw [shift={(223.24,129.29)}, rotate = 0.21] [color={rgb, 255:red, 0; green, 0; blue, 0 }  ][line width=0.75]    (10.93,-4.9) .. controls (6.95,-2.3) and (3.31,-0.67) .. (0,0) .. controls (3.31,0.67) and (6.95,2.3) .. (10.93,4.9)   ;
\draw  [fill={rgb, 255:red, 0; green, 0; blue, 0 }  ,fill opacity=1 ] (357.55,129.79) .. controls (357.55,127.47) and (359.43,125.59) .. (361.74,125.59) .. controls (364.06,125.59) and (365.94,127.47) .. (365.94,129.79) .. controls (365.94,132.11) and (364.06,133.99) .. (361.74,133.99) .. controls (359.43,133.99) and (357.55,132.11) .. (357.55,129.79) -- cycle ;

\draw (268,102.93) node [anchor=north west][inner sep=0.75pt]   [align=left] {system$\displaystyle _{1}$};
\draw (401,102.93) node [anchor=north west][inner sep=0.75pt]   [align=left] {system$\displaystyle _{2}$};
\draw (355,136.24) node [anchor=north west][inner sep=0.75pt]  [font=\small]  {$x_{0}$};
\end{tikzpicture}
\caption{The system on the left (right) of the defect located at $x=x_0$ will be denoted by the superscript $1$ ($2$).}\label{fig:defect_conv}
\end{figure}

In $d=1+1$ integrable classical field theories, an integrable defect is introduced as an internal discontinuity or boundary condition at a fixed point $x_0$ in the spatial direction, see figure \ref{fig:defect_conv}. Although this definition of a defect might seem to compromise integrability from the onset, obviously breaking translation invariance and thus ruining already the conservation of momentum, it is possible to introduce defects whilst preserving integrability.
Indeed, the study of integrable defects in integrable field theories were initiated in the Lagrangian formalism in \cite{Bowcock:2003dr,Bowcock:2004my}, as an extra term in the total Lagrangian. The defect contribution to the total Lagrangian is fixed by demanding it modifies the momentum as to be again conserved. At first sight this approach seems to be only engineered to salvage the conservation of the first charge and nothing in principle guarantees that the defect will preserve sufficiently many conserved charges to warrant for the total system to remain integrable\footnote{In this context, one can distinguish two types of defects. Type I defects do not carry any additional degree freedom and realise defects that only require fields of the original theory. It was realised however, that the definition of type I was not sufficient to realise defects in certain integrable systems. The solution is to allow for the defects to interact through their own degree of freedom (localised at the defect location) \cite{Corrigan:2009vm}. These are called type II or dynamical defects.}. In support of this approach to add defects to integrable field theories was the crucial observation that the defect conditions at the defect location coincide with ``frozen'' B\"acklund transformations\cite{Corrigan:2005iq}. Using this insight, it was shown in  \cite{Caudrelier:2008zz} that the integrable defect indeed preserved integrability by constructing the generating function for the whole tower of modified conserved charges. In addition, this observation gave a new momentum to the study of integrable defects as one can study the B\"acklund transformations for the integrable system to introduce integrable systems defects, see e.g \cite{Gomes:2015hra,Aguirre:2016zrj,Caudrelier:2014gsa}. Although this approach guarantees the conservation of an infinite set of conserved charges (i.e. \textit{weak} integrability) the question of the involution of the charges ( or \textit{strong}/Liouville integrability) is difficult to address due to the divergence in the Poisson brackets at the localised defect, see \cite{Caudrelier:2014oia} for a discussion.

A second, tangent approach, first initiated by \cite{Habibullin:2007ww} and systematically studied in several integrable system in \cite{Avan:2011sr,Doikou:2012fc,Doikou:2016oej}, imposes Liouville integrability from the outset. The idea is to construct a modified monodromy matrix for the total system by gluing the two systems through a defect contribution such that the Sklyanin quadratic algebra remains satisfied. One has in addition to impose consistency relation, dubbed ``sewing/jumping conditions'', between the fields in the bulk theories and those of the defect matrix.  Although this approach warrants for the involution and conservation of the charges, it can clearly only be applied to integrable systems that are ultralocal, or alternatively, admit an ultralocal representation of the Poisson algebra of their Lax connection.  This method was first concretely realised for the non-linear Schr\"odinger model \cite{Avan:2011sr} and later in many other systems, amongst which the Faddeev-Reshetikhin model or ultralocal form of the $SU(2)$ principal chiral model \cite{Doikou:2012fc}.

Let us finally remark that although these two approaches seem diametrically different, both can be in principle used to implement integrable defects in a given integrable system and one should expect them to be somehow related. Indeed, the B\"acklund defects and the Liouville defects were indeed reconciled and clarified their relation in \cite{Caudrelier:2014oia} by using a multisymplectic approach.

In this letter we will make the first steps towards understanding how integrable defects can be introduced in a class of (1+1)-dimensional integrable field theories that feature target spaces with deformed symmetries, known as Yang-Baxter models. In particular, we show how the two methods above have to be modified to describe integrable defects Yang-Baxter models. In section \ref{sec:YBmodels}, we review some basic properties of Yang-Baxter models, setting the stage for the addition of defects in Yang-Baxter models. In section \ref{sec:monodromy_def}, using the modified monodromy matrix approach, we construct a continuous family of Liouville integrable defects for the $S^2$ non-split Yang-Baxter model. In section \ref{sec:baecklund}, we show how B\"acklund transformation can be constructed for the split and non-split Yang-Baxter deformation. Using these we specialise to the defect location and give the form of the defect matrix. We illustrate the construction and provide explicit expressions for the non-split $SU(2)$ Yang-Baxter model.

\section{Yang-Baxter models 101}\label{sec:YBmodels}
 In this section, we will concisely review the Yang-Baxter models and the concepts we need to introduce integrable defects for these models. Extensive discussions on Yang-Baxter models including their relation to generalised dualities can be found in the recent reviews \cite{Thompson:2019ipl,Hoare:2021dix} or the original papers \cite{Klimcik:2002zj,Klimcik:2008eq}.

 \subsubsection*{The principal chiral model} 
For a semisimple Lie group $G$, the principal chiral model (PCM) is described by a map $g:\Sigma \rightarrow G$ with the action
\begin{align}
	S_\mathrm{PCM}= \int \mathrm d^2 \sigma \, \langle g^{-1}\partial_- g,g^{-1}\partial_+ g \rangle\,,
\end{align}
where $\langle -,-\rangle=\mathrm{tr}(-,-)$ is the Killing form on the Lie algebra $\frak g$ of $G$. The global invariance of the action under left and right translation of the principal chiral field $g$ under a constant group element implies the existence of two conserved currents $J_\pm^R =-\partial_\pm gg^{-1}$ and $J_\pm^L =g^{-1}\partial_\pm g$. The equations of motion of the principal chiral model take the form of a flatness and conservation equation for the current
\begin{align}
	\partial_+ J_-^{L/R}+\partial_- J_+^{L/R}=0\,, \quad \partial_+ J_-^{L/R}-\partial_- J_+^{L/R} +[J_+^{L/R},J_-^{L/R}] =0\,.
\end{align}
In what follows we will work exclusively with the left current $J_\pm\equiv J^{L}_\pm$.
Any flat and conserved current capturing the dynamics of a system can be used to construct the Mikhailov and Zakharov Lax connection 
\begin{align}\label{eq:Lax_PCM}
	\mathscr L_\pm(x,t;\xi)=-\frac{J_\pm}{1\pm\xi} \,,
\end{align}
whose flatness is equivalent to the compatibility condition for an auxiliary linear system by the equation
\begin{align}\label{eq:aux_system_PCM}
	\partial_\pm \Psi(x,t;\xi)= - \mathscr L_\pm(x,t;\xi)\Psi(x,t;\xi)\,,
\end{align}
with solutions taking valued in the loop algebra $\mathpzc L \frak g\equiv C^\infty(S^1,\frak g)$.
 \subsubsection*{The Yang-Baxter models}

 Yang-Baxter models are constructed starting from a Lie group $G$ and a linear map $R:\frak g\rightarrow \frak g$ verifying the modified classical Yang-Baxter equation (mCYBE)
\begin{align}\label{eq:mCYBE}
	[RX,RY]-R([RX,Y]+[X,RY])=-c^2[X,Y]\,.
\end{align}
Trivial rescaling of the linear map $R$ leads to three possible values of $c\in \{0,1,i\}$. Solutions corresponding to these three values are called homogeneous, split and non-split, respectively.
The action of the Yang-Baxter model, as an integrable deformation of the principal chiral model (PCM), is described by a map $g:\Sigma \rightarrow M$, for $M$ a Lie group manifold or coset space. When $g$ is valued in a Lie group\footnote{For coset spaces $M$ the action is structurally similar but includes suitable projectors.} that action takes the form \cite{Klimcik:2002zj}
\begin{align}\label{eq:actYB}
	S_\mathrm{YB}= \int \mathrm d^2 \sigma \, \langle g^{-1}\partial_- g, \left(1-	\eta R_g\right)^{-1}g^{-1}\partial_+ g \rangle\,,
\end{align}
where $\Sigma$ is the two-dimensional worldsheet and $R_g=\mathrm{Ad}_{g^{-1}}R\mathrm{Ad}_{g}$.  The existence of a Lax connection in both cases relies heavily on $R$ solving the mCYBE. Historically the non-split Yang-Baxter models are also sometimes called $\eta$-models.

 \subsubsection*{Conserved currents of the Yang=Baxter models} 
 The deformation operator $(1-\eta R)$ in the Yang-Baxter model in eq. \eqref{eq:actYB} breaks the initial global symmetry group $G_L\times G_R$ to $G_L\times U(1)^{\mathrm{rk}(G)}$, hence breaking the right-invariant symmetry. The left current remains unbroken and takes on the form
 \begin{align}\label{eq:conservation_YB_current}
 	\partial_+ J_-^\eta+\partial_- J_+^\eta=0\,, \quad \text{where}\quad J^\eta_\pm=  \frac{1\pm c\eta}{1\pm \eta R_g}g^{-1}_\eta\partial_\pm g_\eta\,.
 \end{align}
 Using the modified Yang-Baxter equation and the equations of motion of the YB model one can show that in fact this current is not only conserved but also flat 
 \begin{align}\label{eq:flatness_YB_current}
 \partial_+ J_-^\eta-\partial_- J_+^\eta +[J_+^\eta,J_-^\eta] =0\,.
\end{align}
We will see in the last section a different, more straightforward argument to the conservation and flatness of the current $J^\eta$.

 \subsubsection*{Yang-Baxter Lax connection}
Using this conserved current \eqref{eq:conservation_YB_current}, one can construct the associated Mikhailov-Zakharov Lax connection which is given by
\begin{align}
	\mathscr L^\eta_\pm = \frac{1}{1\mp \xi}\frac{1\pm c\eta}{1\pm \eta R_g}g^{-1}_\eta\partial_\pm g_\eta=\frac{1}{1\mp \xi} J^\eta_\pm\,.
\end{align}
Flatness of this connection 
\begin{align}
 \partial_+ \mathscr L^\eta_-(\xi)-\partial_- \mathscr L^\eta_+(\xi) +[\mathscr L^\eta_+(\xi),\mathscr L^\eta_-(\xi)] =0\,,
\end{align}
encodes the equations \eqref{eq:conservation_YB_current} and \eqref{eq:flatness_YB_current} summarising the dynamics of the Yang-Baxter model and ensuring the existence of an infinite number of conserved charges.

 \subsubsection*{Some properties of the $R$-matrix} 
Given a solution $R$ of the mCYBE \eqref{eq:mCYBE}, one can show that $\frak g$ admits a second Lie bracket $[-,-]_{ R}=([ R-,-]+[-, R-])$. We will denote this second Lie algebra $\frak g_{ R}=(\frak g,[-,-]_{ R})$, where here $\frak g$ is understood to be the vector space. The couple $(\frak g,\frak g_R)$ is called a bialgebra and they can be embedded together into the so-called Drinfel'd double algebra $\frak d=\frak g\oplus \frak g_R$. It turns out that the solution of the mCYBE $R$ can be used to define an injective homomorphism of $\frak g_{R}$ into $\frak d$ when $R$ is a split or non-split solution. In particular 
	\begin{align}\label{eq:isom_gR}
	\frak g_{R} \hookrightarrow \mathfrak d: 	 X \mapsto \begin{cases}
 	&\gamma ({R}^+X,{R}^-X)\quad \text{(split)}\\
 	&\gamma {R}^\pm X \qquad \qquad \text{(non-split)}
 \end{cases}\,,
	\end{align}
		where $\gamma$ is an arbitrary constant and $R^\pm=R\pm c1$, see e.g. \cite{semenov2008integrable,Vicedo:2015pna} for a detailed discussion.
We will need these maps to extract the Lax connections of the Yang-Baxter-models from the extended solutions of the principal chiral model.

Let us finally briefly summarise the form of the double Lie algebra $\frak d$ in the three types of solutions for the mCYBE \eqref{eq:mCYBE}.
\begin{itemize}
\item Split case ($c=i$)	: The solution is unique and the double $\frak d$ is isomorphic to the complexified algebra $\frak g^\mathbb{C}$. The  algebra $\frak g$ and its dual algebra $\frak g_R$ can be retrieved by identifying $\frak g$ as the real form of $\frak g^\mathbb{C}$ and performing the Iwasawa decomposition $\frak g^\mathbb{C}=\frak g\oplus \frak{an}=\frak g\oplus \frak g_R$. 
\item Non-split case ($c=1$)	: The solution is also unique and the double $\frak d$ is isomorphic to $\frak g\oplus \frak g$. The original algebra $\frak g$ is embedded diagonally in the double $\frak g\cong \frak g^\delta=\{(x,x)\mid x\in \frak g\}\subset \frak d$. To extract the dual algebra we use the isomorphisms \eqref{eq:isom_gR}, $\frak g_R=\{R^+X,R^-X\mid X\in \frak g\}\subset \frak g_\delta$.
\item Homogeneous case ($c=0$): For the homogeneous case, the solution is in general not unique. When restricting to skew-symmetric solutions, one can show that the solutions are in bijective correspondence with so-called quasi-Frobenius Lie subalgebras of $\frak g$ \cite{stolin1991rational,stolin1999rational}. 
\end{itemize}
In this letter, we will consider  split and non-split solutions exclusively.
 
At the level of the group, the decomposition at the level the algebra discussed above might not be true globally. Here however we will assume that the double is simple\footnote{or we can also restrict ourselves to the so called main cell of a double coset decomposition explained in \cite{alekseev1994symplectic}. We also take the $\frak g$ algebras for which the mCYBE is solved to be semi-simple.} such that the we can perform the decomposition $D=GG_R$ at the level of the group unhindered.

 \subsubsection*{Ultralocality and Hamiltonian formulation}
Both the principal chiral model and its integrable deformations suffer from non-ultralocality. That is, the equal-time Poisson-bracket of their Lax connection does not take the Sklyanin form but instead features an extra term containing the derivative of the Dirac-delta distribution
\begin{align}
	\{\mathscr L(x;\xi)\overset{\otimes}{,}\mathscr L(y;\mu)\}=[r(\xi,\mu),\mathscr L(x;\xi)\otimes \mathscr L(y;\mu)]\delta_{xy}+A(x,y;\xi,\mu)\partial_x\delta_{xy}\,,
\end{align}
for some function $A(x,y;\xi,\mu)$, $\delta_{xy}=\delta(x-y)$ and where the $r$-matrix has to be antisymmetric.
However, as was first proposed by Maillet in \cite{Maillet:1985fn,Maillet:1985ek}, for some non-ultralocal systems, amongst which the principal chiral model and the Yang-Baxter models, the Poisson-Bracket can be written in a form generalising the Sklyanin identity and ensures the involution of the conserved charges. The generalisation is achieved by allowing for the $r$-matrix to be non-skew-symmetric. The failure of this new $r$-matrix to be skew-symmetric is measured by the so-called twist function $\varphi(\xi)$ that depends on the spectral parameter $\xi$. 
More explicitly, we have that the $r$-matrix entering the Maillet bracket deforms the canonical $r$-matrix $r^0(\xi,\mu)$ on the loop algebra of $\frak g$:
\begin{align}
	r_{12}(\xi,\mu)=r_{12}^0(\xi,\mu)\varphi^{-1}(\xi)\,.
	\end{align}
In this letter we will not need the explicit form of the Maillet bracket but merely need the expressions for the twist functions.

Sigma-model are generically non-ultralocal and a prime example is the principal chiral model. As an ultralocal model, the Poisson algebra of its Lax connection is thus characterised by its twist function which is given by
  \begin{align}\label{eq:twistfct_PCM}
  	\varphi_\mathrm{PCM}(\xi)= \frac{1-\xi^2}{\xi^2}\,.
  \end{align}
Note already that the twist function features a double pole at the origin. This fact will be important when introducing the dressing method in section \ref{sec:PCM_dressing}. In \cite{Delduc:2013fga}, an Hamiltonian perspective on the deformation of the principal chiral model leading to the Yang-Baxter models was proposed to prove the strong or Hamiltonian integrability of these models. In this picture, and in contrast to the Lagrangian perspective discussed above, the deformation is achieved by keeping the Lax of the principal chiral model unchanged but deforming the twist function. Reality conditions allows for the double pole of the principal chiral model to split symmetrically, either settling on the complex axis of the spectral plane at $\xi=\pm i\eta$ or on the real axis at $\xi=\pm \eta$. In particular the twist function for the split and non-split Yang-Baxter models (up to an overall constant) is given by
   \begin{align}\label{eq:twistfct_YB}
  	\varphi_\mathrm{YB}(\xi)= \frac{1-\xi^2}{\xi^2-c^2\eta^2}\,.
  \end{align}
The pair of poles on the so-called complex branch realise the non-split ($c=i$), while the poles on the real branch lead to the split ($c=1$) Yang-Baxter model. 
For the homogeneous Yang-Baxter models, the effect of the deformation in fact does not change the pole structure of the principal chiral model twist function, but rather reshuffles the expression of the Lax in terms of its defining fields.
  
\section{Liouville integrable defects}\label{sec:monodromy_def}

In this section we first review how integrable defects can be introduced by means of a modified monodromy matrix together with a set of regularisation conditions. We then briefly introduce the sausage model, which is equivalent to the non-split Yang-Baxter model on the two-sphere, and its ultralocal formulation. Having set the ground, we show how Liouville integrable defects can be introduced in the Yang-Baxter model on $S^2$. Notably, the Yang-Baxter model on $S^2$ admits a continuous family of integrable defects.  This feature is not shared by Liouville integrable defects of the undeformed principal chiral. Indeed the presence of an additional parameter can be traced back to the trigonometric r-matrix underlying the Sklyanin identity of the deformed model.

\subsection{Modified monodromy approach to integrable defects}

This method introduces defects by inserting a defect operator in the monodromy, leading to a ``modified monodromy''  \cite{Avan:2011sr}. The challenge however is as we will see that this method assumes ultralocality, ruling out its naive applicability to integrable sigma-models. Denote by $\mathscr L(\xi)$ the Lax connection of a fixed (ultralocal) integrable system whose Sklyanin identity is verified for an $r$-matrix $r(\xi)$. We introduce a spectral-parameter dependent defect operator $\mathcal{D}(x_0,\xi)$ localised at some space point $x_0$ incorporating non-trivial sewing conditions between the bulk theories on either side. This is done by invoking monodromy matrices of systems on both sides of the defect at $x_0$ and glue them together with an Ansatz for the defect matrix $\mathcal D(x_0,\xi)$. Following the notation set in Figure \ref{fig:defect_conv}, the total monodromy matrix reads \cite{Avan:2010xd}
\begin{align}
	\mathcal T(L,-L,\xi)&=T^{(2)}(L,x_0^+,\xi)\mathcal D(x_0,\xi)T^{(1)}(x_0^-,-L,\xi)\notag\\
	&=\mathcal P\exp\left(\int^L_{x_0^+} \mathscr L^{(2)}_x(x)\right)\mathcal D(x_0,\xi)\mathcal P\exp\left(\int^{x_0^-}_{-L}\mathscr L^{(1)}_x(x)\right)\,,
\end{align}
where $x_0^\pm$ are infinitesimally close to the left/right of the defect location $x=x_0$. 
Imposing Liouville integrability on the total system by demanding the modified monodromy to verify the quadratic Poisson algebra or Sklyanin identity
\begin{align*}
	\lbrace \mathcal T(L,-L,\xi)\stackrel{\otimes}{,} \mathcal T(L,-L,\mu )\rbrace =\left[ r(\xi-\mu), \mathcal T(L,-L,\xi)\otimes\mathcal T(L,-L,\mu )\right]\,,
\end{align*}
requires for the defect operator to verify the same quadratic algebra
\begin{align}\label{eq:sklyanin_defect}
	\lbrace \mathcal D(\xi)\stackrel{\otimes}{,} \mathcal D(\mu) \rbrace =\left[ r(\xi-\mu), \mathcal D(\xi)\otimes\mathcal D(\mu)\right]\,,
\end{align}
 where $r(\xi)$ is the same r-matrix as for the bulk theories. In addition, one has to impose sewing conditions across the defect to avoid singular contributions from the zero curvature condition of the Lax connection \cite{Avan:2008xr,Avan:2010xd}. In particular, at the location of the defect the zero curvature condition translates into the condition 
 \begin{align}\label{eq:jumping_condition}
 	\frac{\mathrm d}{\mathrm d t}\mathcal D(x_0)= \tilde{\mathscr{L}}_t^{(2)}(x_0)\mathcal D(x_0)-\mathcal D(x_0)  \tilde{ \mathscr{L}}_t^{(1)}(x_0)\,,
 \end{align}
 where the matrices $\tilde{ \mathscr{L}}_t^{(i)}(x_0)$ are the time components of the Lax pair evaluated at the defect location and are derived by demanding  analyticity at the defect, i.e. $ { \mathscr{L}}_t^{(i)}(x_0^\pm)\rightarrow  \tilde{ \mathscr{L}}_t^{(i)}(x_0)$.  That is the condition \eqref{eq:jumping_condition} effectively describes the jump/sewing over the defect.

\subsection{Liouville integrable defects in the $S^2$  non-split Yang-Baxter model}

The so-called $2d$ ``sausage model'' was first written down in \cite{Fateev:1992tk} as an integrable deformation of the $O(3)$ non-linear sigma-model and shortly after shown to admit a Lax form \cite{Lukyanov:2012zt}. Later, in \cite{Delduc:2013fga,Hoare:2014pna}, it was shown to fit in the Yang-Baxter model family as the non-split Yang-Baxter deformation on the two-sphere $SU(2)/U(1)\cong S^2$. The sausage model is an integrable deformation of the $O(3)$ non-linear sigma-model. Although in this formulation this integrable deformation of the (coset) principal chiral model is non-ultralocal, it is possible to reformulate its dynamics in an ultralocal form. Inspired by the existence of an ultralocal  form of the $O(3)$ model obtained by a ``gauge'' transformation \cite{Bytsko:1994ae}, the authors in \cite{Bazhanov:2017nzh} constructed an ultralocal version of the sausage model. 

We will use the coordinates in terms of the sausage model given in  \cite{Bazhanov:2017nzh}, which we review in this section. That is the $S^2$ Yang-Baxter model or sausage model is given by the Lagrangian 
\begin{align}
	\mathcal L=\frac{1}{2}\frac{\partial_\mu \mathbf n \partial^\mu \mathbf n}{\kappa^{-1}-\kappa n^2_3}\,,
\end{align}
where $\kappa \in (0,1)$ is the deformation parameter and $\mathbf n=(n_1,n_2,n_3)$ is a three-dimensional vector such that $\mathbf n \cdot \mathbf n =1$. Although naively non-ultralocal, a different flat Lax connection can be shown to satisfy an ultralocal Poisson algebra, see \cite{Bazhanov:2017nzh}. The Sklynanin identity is then satisfied for this new Lax connection with the original trigonometric r-matrix 
\begin{align}\label{eq:trig_rmatrix}
	r(\mu)=\frac{1}{\sinh \mu}(\sigma_1\otimes \sigma_1 +\sigma_2\otimes \sigma_2 + \cosh \mu \,\sigma_3\otimes \sigma_3)\,.
\end{align}
In what follows we will refer to this system as the sausage model but with the understanding each time that it is equivalent to the $S^2$ non-split Yang-Baxter model.

Given the ultralocal representation of the 2d sausage model, we propose the following Ansatz for the defect of the $\eta$-deformed $S^2$-model\footnote{This particular form of a matrix verifying versions of the Sklyanin identity and in different contexts has already appeared in \cite{Avan:2010xd,Bazhanov:2017nzh}, what is new here is its interpretation as a defect matrix for the $\eta$-deformed two-sphere.}
\begin{align}\label{eq:Ansatz_defect_sausage}
	\mathcal D(\xi)=\xi \mathds 1 +
\frac{1}{2}\left(
\begin{array}{cc}
 \mathcal{F} (\xi )  \zeta^z&  \mathcal{G}(\xi ) \zeta^-  \\
 \mathcal{G}(\xi ) \zeta^+  & -\mathcal{F}(\xi ) \zeta^z \\
\end{array}
\right)\,,
\end{align}
where the variables $\zeta^i$, upon requiring this defect Ansatz to satisfy the Sklyanin identity, have to generate an $\frak{sl}(2)$ algebra
\begin{align}\label{eq:sl2alg}
	\{\zeta^z,\zeta^\pm\}= \pm 2 \zeta^\pm\,,\quad \{\zeta^+,\zeta^-\}=\zeta^z\,.
\end{align}
and the functions $\mathcal{F}(\mu )$ and $\mathcal{G}(\mu) $ are given by 
\begin{gather}
\begin{aligned}
\mathcal{F}(\mu ) &= \frac{\mu(1+\delta )\mu^{-1}+(1-\delta ) \mu }{(1-\delta ) \mu -(1+\delta )\mu^{-1}}\\
\mathcal{G}(\mu)&= \frac{2 \mu\sqrt{1-\delta ^2}}{(1-\delta ) \mu -(1+\delta )\mu^{-1}}\,,
\end{aligned}
\end{gather}
where $\delta$ is an arbitrary constant.
It can be readily checked to verify the Sklyanin identity \eqref{eq:sklyanin_defect} for the trigonometric $r$-matrix given in \eqref{eq:trig_rmatrix}. Let us note that in comparison to the defect Ansatz proposed in \cite{Doikou:2012fc} for the (undeformed) $SU(2)$ principal chiral, the sausage model admits a continuous family of defects parametrised by the parameter $\delta$. This freedom is a direct consequence of the sausage model being described by a trigonometric rather than rational r-matrix. Conversely the defect Ansatz \eqref{eq:Ansatz_defect_sausage} does not solve the Sklyanin identity for the rational r-matrix considered in \cite{Doikou:2012fc} for the undeformed $SU(2)$ principal chiral model. Finally let us remark that the parameter $\delta$ that enters the defect Ansatz cannot be identified with the deformation parameter $\kappa$, as one can check that this identification is incompatible with the sewing conditions, that we now discuss.

Satisfying only the Sklyanin identity  \eqref{eq:sklyanin_defect} for the defect is not sufficient. Integrability is only preserved provided the additional sewing or jumping conditions can be solved for an adequate form of the time component Lax at the location of the defect given in \eqref{eq:jumping_condition}. In order to do so, first note that in general $\mathscr{L}_t$ is given in terms of the monodromy matrix by \cite{Avan:2011sr,Faddeev:2007} 
\begin{equation}\label{eq:V from monodromy}
\mathscr{L}_t(x,\xi,\mu)=t^{-1}(\xi) \mathrm{tr}_a \left(\mathcal{T}_a(L,x,\xi) r_{ab}(\xi,\mu)  \mathcal{T}_a(x,-L,\xi) \right)\,.
\end{equation}
The subscript $a$ is a tensor product space index, such that $\mathcal{T}_a, \mathcal{D}_a$ are given as $\mathcal{T}\otimes \mathds{1}$ and $\mathcal{D} \otimes \mathds{1}$ and $t(\lambda)=\mathrm{tr}(\mathcal{T}(L,-L,\lambda) )=\mathrm{tr}(T^{(2)}(L,x_0,\lambda)\mathcal{D}(x_0,\lambda)T^{(1)}(x_0,-L,\lambda))$. 
Depending on whether the above expression is evaluated on the bulk or at the defect location $x_0$ will lead to a generically different results. Looking at the right bulk theory and its limit towards the defect location, we have that 
\begin{gather}
\begin{aligned}
\mathscr{L}_t^{(2)}(x,\xi,\mu)& = \frac{t^{-1}(\xi)}{\xi/\mu-(\xi/\mu)^{-1}}\left(M_o + \frac{\xi/\mu-(\xi/\mu)^{-1}}{4} (M_d - \det(M_d)M_d^{-1} \right)\,,\\
\tilde{\mathscr{L}_t}^{(2)}(x,\xi,\mu)& =  \frac{t^{-1}(\xi)}{\xi/\mu-(\xi/\mu)^{-1}}\left(N_o + \frac{\xi/\mu-(\xi/\mu)^{-1}}{4} (N_d - \det(N_d)N_d^{-1} \right)\,.
\end{aligned}
\end{gather}
where  $M_d$ and $M_o$ denote, respectively, the diagonal and off-diagonal parts of $M$ the matrices $M$ and $N$ given by 
\begin{align*}
M&=\ T_a^{(2)}(x,x_0,\xi) \mathcal{D}_a( x_0,\xi)T_a^{(1)}(x_0,-L,\xi)T_a^{(2)}(L,x,\xi)\,,\\
N&=\mathcal{D}_a( x_0,\xi) \ T_a^{(1)}(x_0,-L,\xi)T_a^{(2)}(L,x_0,\xi)\,.
\end{align*}
The sewing conditions are obtained by imposing that the bulk fields glue continuously along the defect, i.e 
\begin{equation}\label{eq:sewing condition}
\lim_{x \to x_0} \mathscr{L}_t^{(2)}(x) = \tilde{\mathscr{L}_t}^{(2)}(x_0)\,.
\end{equation}
This will lead to the sewing condition which will involve the bulk fields as well as the $\frak{sl}(2)$-variables $\zeta^i$ given in \eqref{eq:sl2alg} that are associated to the defect itself. One has similar relations for time-component of the left bulk and defect Lax connections  $\mathscr{L}_t^{(1)},\tilde{\mathscr{L}_t}^{(1)}$.

Calculating the sewing conditions from the gluing condition in eq. (\ref{eq:sewing condition}) is carried out by expanding the monodromy matrix order by order in powers of the spectral parameter. 
A generic expansion of the monodromy is given by  \cite{Faddeev:2007}
\begin{equation}\label{eq:monodromy_expansion}
T^{(1/2)}(x,y,\xi) = (\mathds 1+W^{(1/2)}(x))e^{Z^{(1/2)}(x,y)}(\mathds 1+W^{(1/2)}(y))^{-1}\,,
\end{equation}
where $W$ and $Z$ are, respectively, anti-diagonal and diagonal matrices that are analytic in the spectral parameter $\xi$ around the poles of the Lax connection
\begin{align*}
W(\xi)&=\sum_{n=0}^{\infty} W_n (\xi -\xi_p)^n\,, \qquad W_n=\left(
\begin{array}{cc}
0 & u_n\\
v_n  & 0 \\
\end{array}
\right)\,,\\
Z(\xi)&=\sum_{n=-1}^{\infty}Z_n(\xi-\xi_p)^n\,, \qquad Z_n=\left(
\begin{array}{cc}
 q_n &0\\
0& r_n  \\
\end{array}
\right)\,,
\end{align*}
where $\xi_p$ is a fixed pole of the Lax connection and the matrix elements $q_n,r_n,u_n,v_n$ are yet to be determined\footnote{In fact as we will see, to zeroth order, the matrix elements from the $Z$ matrix will not play a role and consequently won't enter the sewing condition.}. Using this form of the matrices $W(\xi)$ and $Z(\xi)$, we get sewing conditions relating their various matrix elements and the fields $\zeta^i$ featured in the defect Ansatz. The Lax connection of the ultralocal sausage model has a pole at $\xi_p=\frac{\sqrt{1+\kappa}}{\sqrt{1-\kappa }}$, expanding the monodromy matrix as in eq. \eqref{eq:monodromy_expansion} around $\xi_p$ to zeroth order: 
\begin{align}
[\mathscr{L}_t^{(2)}]_0&=\frac{1}{\mathfrak{J_-}(\xi)}
\left(
\begin{array}{cc}
 \frac{1}{4}\mathfrak{J}_+(\xi) &
   - \xi\,  u^{(2)}_0\\
  \xi \,v^{(2)}_0 &
  -  \frac{1}{4}\mathfrak{J}_+(\xi)\\
\end{array}
\right)\,,
\end{align}
for the expression on the bulk. The subscript 0 denotes the zeroth order contribution and the overall denominator is given by
\begin{align}
	\mathfrak{J}_\pm(\xi)= \kappa_\pm(\xi)\left(1\pm u^{(2)}_0 v^{(2)}_0\right)\,,\;\; \text{with}\;\; \kappa_\pm(\xi)=\xi_p(1\pm (1-\kappa)\xi^2)\,.
\end{align}
At the defect location we get
\begin{align}\label{eq:t_Lax_2_def_loc}
[\tilde{\mathscr{L}_t}^{(2)}]_0=\frac{1}{\tilde{\mathfrak{J}}_-(\xi)}
\left(
\begin{array}{cc}
 \frac{1}{4}\tilde{\mathfrak{J}}{}_+(\xi) & -
 \left(\Omega_--Z^- v^{(1)}_0\right) \xi u^{(2)}_0 \\
\left( \Omega_+v^{(1)}_0-Z ^+ \right)\xi& - \frac{1}{4}\tilde{\mathfrak{J}}{}_+(\xi) \\
\end{array}
\right)
\end{align}
where $\Omega_\pm=2(\delta- \kappa) \pm (1-\delta  \kappa)  \zeta ^z$ and $Z^\pm =\sqrt{1-\delta ^2}  \sqrt{1-\kappa ^2} \zeta ^\pm$. The overall scaling factor in \eqref{eq:t_Lax_2_def_loc} is given by
\begin{align}
\tilde{\mathfrak{J}}_\pm(\xi) &= \kappa_\pm(\xi) \left( \Omega_--Z^-v_0^{(1)}-(\Omega_+v_0^{(1)}-Z^+)u_0^{(2)}\right) 
\end{align}
Imposing the gluing or jumping constraint in eq. (\ref{eq:sewing condition}) for the time-component of the Lax connection right of the defect $\mathscr L_t^{(2)}$ and at the defect $\tilde{\mathscr{L}}_t^{(2)}$ we get the following sewing condition 
\begin{align}\label{eq:sewing bulk 2}
v^{(1)}_0 v^{(2)}_0 Z ^-&=Z ^++ \Omega_+v^{(1)}_0 - \Omega_-v^{(2)}_0\,.
\end{align}
 Performing the same computation for the fields on the system left of the defect, we get 
\begin{align}\label{eq:sewing bulk 1}
u^{(1)}_0 u^{(2)}_0 Z^+ &=Z ^-- \Omega_+u^{(1)}_0 + \Omega_-u^{(2)}_0\,.
\end{align}
We see that the two conditions are related by exchanging
$Z^+ \leftrightarrow Z^-$ (or equivalently $\zeta^+ \leftrightarrow \zeta^-$) and 
$v^{(i)}_0 \leftrightarrow -u^{(i)}_0$. A similar symmetry was already noticed in \cite{Doikou:2012fc} for the ultralocal $SU(2)$ principal chiral model.

The defect \eqref{eq:Ansatz_defect_sausage} for the $S^2$ non-split Yang-Baxter model features two continuous parameters: the deformation parameter $\kappa$ of the sausage model and the arbitrary constant $\delta$ carried by the defect itself. The latter is a novel feature of the deformed model in comparison with the defect for the undeformed $SU(2)$ principal chiral mdoel presented in \cite{Doikou:2012fc}. Finally, using the expansion of the monodromy matrix given in eq. \eqref{eq:monodromy_expansion} together with the defect matrix, one can extract the defect contribution to the lowest modified conserved charge
\begin{align}
\mathfrak D_0 &= \frac{\xi_p \left( \Omega_--Z^-v_0^{(1)}-(\Omega_+v_0^{(1)}-Z^+)u_0^{(2)}\right)}{2(\kappa-\delta)(1-u_0^{(2)}v_0^{(2)})} \,.
\end{align}
Using the sewing condition at lowest order in eq. \eqref{eq:sewing bulk 2} simplifies the corresponding defect contribution to
\begin{align}
\mathfrak D_0 &= \frac{\xi_p\left( Z^+v_0^{(2)} -\Omega_+ \right)}{2(\kappa-\delta)}  \,.
\end{align}
The derivation of the higher order charged can then performed in a similar way, each time computing at the same time the sewing condition of the corresponding order.

\section{B\"acklund transformations for inhomogeneous Yang-Baxter models}\label{sec:baecklund}
 A B\"acklund transformation\footnote{One can differentiate between auto-B\"acklund and B\"acklund transformations, where the suffix ``auto'' refers to the case where the differential system relates two solutions of the \textit{same} non-linear differential equation. Since we will only consider auto-B\"acklund transformations, relating solutions of the equations of motion of the Yang-Baxter model, we will just write ``B\"acklund'' throughout.} is a system of differential equations relating two different solutions of a given non-linear differential equation describing e.g. the dynamics of a physical system or the defining relations of certain surfaces. Strikingly, B\"acklund transformations appear to be closely related to integrable systems as the differential equations that admit such transformations often coincide with that of an integrable system, see \cite{rogers2002backlund} for an extensive review. The main application of B\"acklund transformations is as a solution generating technique. Indeed, given a solution, often taken to be the vacuum or trivial solution, the B\"acklund transformation yields a differential equation that is often linear or of a lower order than the equations of motion.  In the context of integrable defects, the defect condition coincide with a B\"acklund transformation ``frozen'' at the defect location, thus relating the two solution separated by a defect \cite{Corrigan:2005iq}. The fields\footnote{Note that, to avoid a concatenation of super- and subscripts, we will in what follows change the notation slightly compared to the previous section. The fields in the system on the left/right of the defect will be denote without/with a tilde instead of a subscript (1)/(2).} $g$ and $\tilde g$ on the left and the right of the system are then related at the location of the defect $x=x_0$ by a defect matrix $K(x_0,t)$, that is
 \begin{align}\label{eq:defect_matrix}
 	\tilde g(x_0,t)=K(x_0,t) g(x_0,t)\,,
 \end{align}
 where $K(x_0,t)$ is the defect matrix that is related to the particular B\"acklund transformation.

 In this section we will first review the dressing method and the related B\"acklund transformations in the principal chiral model, setting the ground for deriving defects in the Yang-Baxter model. To construct the latter, we need to adopt the Hamiltonian approach to integrable deformations: the deformation is seen as changing the pole structure of the twist function and thus the Poisson brackets of the field in the cotangent space. The Yang-Baxter fields $g,\tilde g:\Sigma\rightarrow G$ on both side of the defect, as well as the defect matrix itself, are then constructed by first evaluating a dressed solution of the principal chiral model at the poles of the twist function and a subsequent group decomposition.
 
 Let us also mention that although defects have never been studied in Yang-Baxter models before, in \cite{Driezen:2018glg}  integrable boundary configuration on the worldsheet, and the corresponding D-branes on the target space, were derived for the related $\lambda$-deformation and then mapped via Poisson-Lie T-duality to the non-split Yang-Baxter model for $G=SU(2)$.
\subsection{Dressing method and B\"acklund transformations in the PCM}\label{sec:PCM_dressing}
It was first observed in \cite{zakharov1978relativistically} and \cite{shabat1972exact} that, using the auxiliary problem one can obtain new solution of the auxiliary problem in eq. \eqref{eq:auxproblem} by ``dressing'' a known solution.  The starting point is the auxiliary problem 
\begin{align}\label{eq:auxproblem}
	\partial_\pm\Psi(x,t;\xi)=-\mathscr L_\pm(x,t;\xi) \Psi(x,t;\xi)\,.
\end{align}
that encodes the flatness condition of the Lax connection as its compatibility condition. Considering the auxiliary problem for the principal chiral model, that is for the Lax connection given in \eqref{eq:Lax_PCM}, note that the principal chiral field can be extracted from the extended solution by evaluating at $\xi=0$, indeed $\Psi(x,t;0)=g$. Denoting a known solution by $\Psi$ and the new putative solution by $\tilde \Psi$, the dressing Ansatz takes the form
\begin{align}\label{eq:dressing_PCM}
	\tilde \Psi(x,t;\xi)=\chi(x,t;\xi)\Psi(x,t;\xi)\,,
\end{align}
where $\chi(x,t;\xi)$ is a spectral parameter dependent function called the dressing factor which particular form is derived by demanding $\tilde \Psi(x,t;\xi)$ to remain a solution of the auxiliary problem and depends also on the specific target space. The chiral field of the dressed solution can then be extracted from the dressed solution by evaluating the relation \eqref{eq:dressing_PCM} at $\xi=0$, i.e. $\tilde g=\chi(0)g$. The dressing factor $\chi(\xi)$ and its inverse are required to be meromorphic functions in the spectral parameter \cite{Ogielski:1979hv,harnad1984backlund} 
\begin{align}
	\chi(\xi)=\mathds 1+\sum_{i=1}^M\frac{Q_i}{\xi-\xi_i}\,,\qquad  \chi(\xi)^{-1}=\mathds 1+\sum_{i=1}^M\frac{R_i}{\xi-\xi_i}\,,
\end{align}
featuring a finite set of simple poles $\{\xi_i\}_{i=1}^K$ and $\{\xi_i\}_{i=1}^K$ where $K$ is an integer. 

The matrices $R_i$ and $Q_i$ are functions only of the worldsheet coordinates and not the spectral parameter. The dressing function is normalised by requiring $\chi(\infty)=\chi^{-1}(\infty)=I$.
Additional analyticity constraint on the residues $Q_i$ and $R_i$ leads to a derivation of their general form, see \cite{harnad1984backlund}. More details on the minimum number of poles and their reality conditions  depend on the target space and can be found in that reference.

Plugging the dressed wave function in eq. \eqref{eq:dressing_PCM} back into the auxiliary system \eqref{eq:aux_system_PCM}, demanding the system to remain invariant and for the currents \eqref{eq:Lax_PCM} featured in the Lax connection to remain independent of the spectral parameter, the currents have to transform as 
\begin{align}\label{eq:pre-Baecklund_PCM}
	\tilde J_\pm= \chi^{-1}(\xi)J_\pm\chi(\xi)+(1\mp\xi)\partial_\pm\chi^{-1}(\xi)\chi(\xi)\,.
\end{align} 
Subsequently evaluating equation \eqref{eq:pre-Baecklund_PCM} at the value $\xi=0$ of the spectral parameter, yields an equation in the algebra $\frak g$ relating the currents of the seed solution $g$ and the dressed solution $\tilde g$
\begin{align}
	\tilde J_\pm= \chi^{-1}(0)J_\pm\chi(0)+\partial_\pm\chi^{-1}(0)\chi(0)\,.
\end{align}
This relation is known as the B\"acklund transformation for the principal chiral model, transforming the current of the principal chiral field of the seed solution $g:\Sigma\rightarrow G$ and that of the dressed solution $\tilde g:\Sigma\rightarrow G$. In the following section we show how, by lifting the construction to the Drinfel'd double and considering an ``auxiliary principal chiral model'' there, one can construct B\"acklund transformations for the split and non-split Yang-Baxter models.

\subsection{B\"acklund transformations for Yang-Baxter models}

Our strategy to identify B\"acklund transformations and thus defect matrices for the Yang-Baxter model will be to use the Hamiltonian approach summarised at the end of section \ref{sec:YBmodels} rather than the Lagrangian approach. In the Hamiltonian setting the starting point are the undeformed principal chiral model Lax connection and its twist function that encodes the non-ultralocal form of its Poisson algebra. The deformed model in the split and non-split case is then obtained by moving the double poles of the Lax matrix whilst preserving certain reality conditions. 
  
As was first observed in \cite{Klimcik:2008eq,Klimcik:2014bta} for the non-split case, the Yang-Baxter field $g:\Sigma\rightarrow G$ can by obtained by evaluating the extended solution of the auxiliary problem of the \textit{principal chiral model} at $\xi=\pm c\eta$, rather than $\xi=0$, and subsequently performing an Iwasawa decomposition $G^\mathbb{C}=AN\cdot G$. The Yang-Baxter field $g_\eta:\Sigma \rightarrow G$ is then identified with the contribution from the compact real subgroup of $G^\mathbb{C}$, i.e.
\begin{align}
G^\mathbb{C} \ni \Psi (\pm c\eta)=b_\eta g_{\eta}\in  AN\cdot G\,.
\end{align}
Later this very observation was explained and generalised to solutions $c=1$ of the mCYBE by using the Hamiltonian formulation of the Yang-Baxter models. The value $\xi=-i\eta$ used in \cite{Klimcik:2014bta}, or more generally $\xi=\pm c\eta$, was identified as the pole of the twist function controlling the deformation \cite{Delduc:2013fga}, see also \cite{Vicedo:2015pna}. In the split case, the wave function takes values in the real double and we have the corresponding decomposition $\Psi(\pm \eta)=g^\delta  b\in G^\delta\cdot G_R=G\times G$. Note that this procedure is consistent with the observation made in the previous section: evaluating the extended solution at the location of a pole of the twist function yields the principal chiral field $g_0$ or Yang-Baxter field $g_\eta$, except that in the latter one has to perform an additional group decomposition.  We will briefly review this argument, as it will be critical to define B\"acklund transformations for Yang-Baxter models.

Starting from the auxiliary problem 
 \begin{align}\label{eq:aux_problem_PCM_LCG}
	\mathscr L_\pm^\mathrm{PCM}  (\pm c\eta)\equiv - \frac{g_0^{-1}\partial_\pm g_0}{1 \pm c\eta}&= -\Psi ^{-1}(\pm c\eta)\partial_\pm \Psi (\pm c\eta)\,,
\end{align}
we can consider this equation now for $\xi$ taking values in $\mathbb C$. In particular, we can now evaluate the wave function $\Psi (\xi)$ at both the real or complex values $\xi=\pm c\eta$ corresponding to the poles of the twist functions \eqref{eq:twistfct_YB} of the split and non-split Yang-Baxter models. Without loss of generality and to avoid a profusion of signs, in what follows  we will fix the sign to $\xi=+c\eta$.  Remarkably, after performing the decomposition, the conserved current of the principal chiral model $J^0_\pm$ turns out to be proportional to the unbroken current of the Yang-Baxter model
 \begin{align}\label{eq:idfct_Jeta_PCMcurrent}
 		\frac{g_0^{-1}\partial_\pm g_0}{1\mp c\eta}= \frac{1\pm c\eta}{1\pm \eta R_g}g^{-1}_\eta\partial_\pm g_\eta= J_\pm^\eta\,,
\end{align}
where the derivation uses that {\color{black} $R\pm c:\frak g_R\hookrightarrow \frak d$} is an injective homomorphism \eqref{eq:isom_gR}, and thus that in particular for any $b\in AN$ one can write $b^{-1}\partial_\pm b=\eta (R- c )X_\pm$ for some $X_\pm\in \frak g$ and the factor $\eta$ is there for convenience. Note that the identity \eqref{eq:idfct_Jeta_PCMcurrent} provides direct proof that the current $J^\eta$ is both conserved and flat, since $J^0_\pm=g_0^{-1}\partial_\pm g_0\propto J^\eta_\pm$. This was exactly the observation mentioned earlier that was used in \cite{Klimcik:2014bta} to re-derive the Lax connection for the (non-split) Yang-Baxter model and construct that of the bi-Yang-Baxter model.

This suggests that we have to consider dressing solutions for the auxiliary system \eqref{eq:aux_problem_PCM_LCG} for the PCM with fields taking values in the double of the split or non split solution. By evaluating at the poles of the Yang-Baxter twist function we will then obtain B\"acklund transformation for the deformed systems. For a seed solution $\Psi $ of the PCM consider the dressed solution for a particular allowed dressing factor $\chi(\xi)$:
\begin{align}\label{eq:dress_Psi}
	\widetilde \Psi (\xi)=\chi(\xi)\Psi (\xi)\,.
\end{align}
In the last section we haver reviewed how the respective currents are related through eq. \eqref{eq:pre-Baecklund_PCM} valued in loop group $\mathpzc L\frak g^\mathbb{C}$. However, now we can instead evaluate this equation at the location of the poles of the deformed twist function $\xi=c\eta$ and thus find a relation between the currents of the Yang-Baxter model
\begin{align}
	\widetilde J_\pm^\eta=\chi(c\eta)^{-1} J_\pm^\eta\chi(c\eta)+(1\mp c\eta)\chi(c\eta)^{-1}\partial_\pm \chi(c\eta) \,.
\end{align}
We can now both extract the Yang-Baxter fields $g_\eta, \tilde g_\eta:\Sigma\rightarrow G$ and interpret the transition between the two fields as crossing a defect as in \eqref{eq:defect_matrix}.
As explained above, the group element $\tilde g_\eta$ can be obtained by performing an Iwasawa decomposition of the right hand side of
\begin{align}
	\tilde\Psi(c\eta)=\tilde b_\eta \tilde g_\eta= \chi(c\eta)\Psi(c\eta)\,.
\end{align}
In addition, after some reshuffling, we can identify the defect matrix $K$ as a `dressed' dressing factor evaluated at the poles of the twist function
\begin{align}\label{eq:defect_YB}
	\tilde g_\eta = \Big(\tilde b_\eta^{-1}\chi(c\eta)b_\eta\Big) g_\eta\equiv Kg_\eta\,.
\end{align}

We will close this section with a couple of comments. Let us first point out that the B\"acklund transformations are interesting in se and not only as an instrument for constructing defects. B\"acklund transformations offer an alternative way to the monodromy matrix expansion to generate the tower of conserved charges underlying the integrable property of the model. Given a periodic solution of the B\"acklund equations, one can construct an infinite number of local conserved currents of the seed solution  \cite{Pohlmeyer:1975nb}. Systematic methods to extract the charges using this method for the PCM were first developed perturbatively in \cite{Ogielski:1979hv} and later a non-perturbative method was presented in \cite{harnad1984backlund}. The relation between the charges generated from the monodromy matrix and the B\"acklund transformations was studied in \cite{Arutyunov:2005nk} in the particular context of strings propagating in AdS$_5\times S^5$. 

Although we have shown here how B\"acklund transformations can be realised for Yang-Baxter models, the dressing method itself for Yang-Baxter models does not seem to straightforwardly follow from the results presented here. This generalisation of dressing method would require one to solve the auxiliary problem with the Yang-Baxter Lax connection together with a way to extract the Yang-Baxter field directly. Here we obtained the B\"acklund transformations for the Yang-Baxter-models by starting from the extended problem for the principal chiral model and subsequently reduced its dressed solution to a solution of the Yang-Baxter-model. Note however that the dressing method has already been considered in the context of the $\lambda$-deformation in \cite{Appadu:2017xku}.

\subsection{Defect matrix for the $SU(2)$ non-split Yang-Baxter model}
In this section we provide an example of two Yang-Baxter fields valued in $SU(2)$ that can be separated by a defect and give the corresponding defect matrix $K$.  We thus need to consider the auxiliary problem of the PCM for $SU(2)^\mathbb{C}=SL(2,\mathbb{C})$. The procedure starts from a vacuum solution, i.e. a principal chiral field $g:\Sigma\rightarrow SL(2,\mathbb{C})$ such that 
\begin{align}\label{eq:vacuum_eqs}
	\partial_-J_{0,+}=\partial_+J_{0,-}=[J_{0,+},J_{0,-}]=0\,.
\end{align}
The auxiliary problem for a vacuum solution is easily integred, yielding the solution $\Psi_0(\xi)$ to the auxiliary problem.
To dress this solution, we consider the simplest possible dressing factor, which for $SL(2,\mathbb{C})$ takes the form \cite{harnad1984quadratic}
\begin{align}\label{eq:dressing_factor}
	\chi(\xi)=\left(\mathds 1+\frac{\mu_0-\xi_0}{\xi-\xi_0}P \right)\,,
\end{align}
where $\mu_0$ and $\xi_0$ are constants taken to be valued  on the unit circle and $P$ is a projector. The latter is computed by choosing two arbitrary 2-dimensional vectors $m$ and $n$ such that $n^\dagger m$ is non-singular, in terms of which it is computed via
\begin{align}
	P=\frac{\Psi_0(\mu_0)m\,n^\dagger\Psi_0(\xi_0)^{-1}}{n^\dagger \Psi_0(\xi_0)^{-1}\Psi(\mu_0)m}\,.
\end{align}
One must be careful, as the dressing factor is not necessarily on element of $SL(2,\mathbb{C})$ and thus neither \`a priori the dressed solution $\Psi(\xi)=\chi(\xi)\Psi_0(\xi)$. To mend this, one adds a normalisation factor to \eqref{eq:dressing_factor} ensuring the dressing factor has unit determinant\footnote{ As was remarked in \cite{harnad1984quadratic}, the dressing factor for the PCM auxiliary problem in $SL(2,\mathbb C)$ (or more generally $GL(2,\mathbb C)$) might feature singularities producing a singular solution from a non-singular seed solution. Although nothing points to these solution being physical not sound, we will abstain from considering any singular dressing factors.}.

\noindent
Consider the following $SL(2,\mathbb C)$ ``vacuum solution'' 
\begin{align}
g_0=\left(
\begin{array}{cc}
 1 & x_- +x_+ \\
 0  & 1 \\
\end{array}
\right)\,, \quad \text{with} \;\,J_{0,\pm}= \left(
\begin{array}{cc}
 0 & 1 \\
 0 & 0 \\
\end{array}
\right)\,,
\end{align}
where we have denoted the light-cone coordinates by $x_\pm=(t\pm ix)/2$. The extended solution can be readily integrated, yielding 
\begin{align}\label{eq:example_psi_vac}
\Psi_0(\xi)=\left(
\begin{array}{cc}
 1 &\frac{x_+}{1-\xi}+ \frac{x_- }{1+\xi} \\
 0 & 1 \\
\end{array}
\right)\,.
\end{align}
To build the dressing factor \eqref{eq:dressing_factor}, we chose two equal vectors $m=n=(1,0)^t$ for which the projector takes the form
\begin{align}
	P= \left(
\begin{array}{cc}
 0 & \frac{x_+ }{1-\mu_0}+\frac{x_- }{1+\mu_0} \\
 0 & 1 \\
\end{array}
\right)\,.
\end{align}
We will take for simplicity $\mu_0=\bar{\lambda}_0=\exp(i\gamma)$ for $\gamma$ an (real) angle. Then the normalisation factor $\chi(\xi)\rightarrow \alpha^{-1}\chi(\xi)$ ensuring that the dressed solution still has unit determinant is
\begin{align}
	\alpha= \sqrt{1+\frac{1-e^{2i\gamma}}{1-ie^{i\gamma}\eta}}\,.
\end{align}
Computing with this information the dressed extended solution \eqref{eq:dressing_PCM} with the dressing function \eqref{eq:dressing_factor} evaluated at the pole of the twist function $\xi=i\eta$ yields
\begin{align}\label{eq:example_psi_dress}
\Psi(\xi)=\left(
\begin{array}{cc}
\alpha^{-1} &\frac{  \left(t-\eta x)(t+i e^{i \gamma } x\right) }{\left(1+\eta ^2\right) \left(1- i\eta e^{i\gamma}\right)}\alpha^{-1} \\
 0 & \alpha \\
\end{array}
\right) = \left(
\begin{array}{cc}
\alpha^{-1} &\alpha^{-1}\beta  \\
 0 & \alpha \\
\end{array}
\right)  \,.
\end{align}
We can now identify the corresponding Yang-Baxter fields $g_\eta, \tilde g_\eta$ and interpret them as two systems separated by a defect. Decomposing the vacuum solution \eqref{eq:example_psi_vac} at $\xi=i\eta$, we get for the Yang-Baxter map of the system left from the defect  
\begin{align}
	g_\eta= \rho^{-1} \left(
\begin{array}{cc}
 1+\eta ^2 & (t-\eta  x) \\
 - (t-\eta  x) & 1+\eta ^2 \\
\end{array}
\right)\,,\quad \rho^2=\left(1+\eta ^2\right)^2+ (t-\eta  x)^2\,.
\end{align}
and from \eqref{eq:example_psi_dress}, the Iwasawa decomposition yields the map characterising the system on the right
\begin{align}
	\tilde g_\eta=\frac{1}{\sqrt{1+|\beta|^2}} \left(
\begin{array}{cc}
\left(\alpha/\bar\alpha\right)^{1/2} & \bar \beta \left(\alpha/\bar\alpha\right)^{1/2} \\
 - \beta \left(\alpha/\bar\alpha\right)^{-1/2} &\left(\alpha/\bar\alpha\right)^{-1/2}  \\
\end{array}
\right) \,,
\end{align}
where $d$ has been defined in \eqref{eq:example_psi_dress}.
Finally we can interpret these two Yang-Baxter fields as being divided by a defect captured by  the defect matrix $K=\tilde b_\eta^{-1} \chi(i\eta)b_\eta$  given by
\begin{align}
	K=  \frac{1}{\sqrt{1+|\beta|^2} \rho}\left(
\begin{array}{cc}
\frac{(1+|\beta|^2)(\rho^2+(1+\eta^2)^2\beta)}{(1+\eta^2)\alpha |\alpha|}+\frac{\alpha^4(t-\eta x)\bar \beta}{ \alpha |\alpha|}&\frac{\beta(1+|\beta|^2)}{\alpha |\alpha|}\frac{(1+\eta^2)^2}{t-x\eta}+\frac{\bar \beta \alpha^2}{|\alpha|} (1+\eta^2) \\
(t-x\eta)|\alpha|& (1+\eta^2)|\alpha|\\
\end{array}
\right)\,.
\end{align}

\section{Conclusions and outlook}

In this short letter, we made the first steps towards implementing integrable defects in a class of  integrable deformations known as Yang-Baxter models. We first considered Liouville integrable defects by modifying the monodromy matrix to include the contribution of a defect matrix. We applied this method to the non-ultra local representation of the $S^2$ non-split Yang-Baxter model. A major hurdle for the implementation of Liouville integrable defects is the non-ultra locality of the Yang-Baxter models. This method has as starting assumptions that the monodromy matrices of the systems on both side of the defect satisfy the Sklyanin identity. To apply this method one has to only consider non-ultralocal models or instead rely on the existence of an ultralocal representation, if it even exists\footnote{One could wonder if the non-ultra locality of the family of principal chiral models is a fundamental property with a few ``happy'' coincides for which we can write down an ultralocal representation or rather that it is the product of a bad choice of ``coordinates'' and that all integrable sigma-models are in fact not fundamentally non-ultralocal.}. As such, the possible application of this method are, due to the absence of a general ultralocal formulation of Yang-Baxter models, quite constrained and we have presented the defect matrix for the only Yang-Baxter model that to our knowledge is known to admit an explicit ultralocal representation. Recent developments, seem to suggest that the ultralocality of non-linear sigma-models might be more of technical, rather than a fundamental, obstacle. For symmetric space sigma models for example, the original regularisation procedure of Faddeev and Reshetikhin can be generalised and one can alleviate the non-ultralocality of these models \cite{Delduc:2012qb}. One could also start from the 4d Chern-Simons approach \cite{Costello:2019tri,Delduc:2019whp} to integrable field theories that encodes both ultralocal and non-ultralocal indiscriminately.

Using a different approach to integrable defect, we showed how B\"acklund transformations for the split and non-split Yang-Baxter models can be obtained by using the Hamiltonian perspective on integrable deformation and using as starting point the auxiliary problem of the principal chiral model.  Following the interpretation of a defect as a ``frozen'' B\"acklund transformation of the defect we give the expression for the defect matrix relating two Yang-Baxter fields. As an example we constructed two Yang-Baxter fields for a deformed $SU(2)$ target space that can be separated by an integrable defect and write down the associated  defect matrix.

The B\"acklund approach can be applied to a much wider range of Yang-Baxter models compared to the modified monodromy method  as it does not rely on the systems being ultralocal. Here the challenge would be to generalise the derived B\"acklund transformation to the Yang-Baxter fields on deformed coset spaces, extending their applicability to holographic backgrounds as well as enabling us to construct B\"acklund defects in the $S^2$ non-split Yang-Baxter model. An obvious strategy in that direction would be to try to generalised the dressing method to Yang-Baxter model directly rather than rely on the dressing method of the principal chiral model and a subsequent reduction as discussed in section \ref{sec:baecklund}. The relevance of such a generalisation of the dressing method to Yang-Baxter models is manifold. It would e.g. give us a direct window towards exploring giant magnons \cite{Spradlin:2006wk,Kalousios:2006xy,Appadu:2017xku} in Yang-Baxter deformed target spaces and thus offer a new direction towards understanding the dual gauge theories via the spectrum of the corresponding integrable spin chain. In addition, a dressing method for Yang-Baxter models could provide a way to derive, from first principles, the uniton solutions for the $SU(2)$ non-split Yang-Baxter model found in \cite{Demulder:2016mja}. We plan to report on this particular direction in the future.

\acknowledgments
SD would like to thank the organisers of the conference and school ``Duality, integrability and deformations 2021'' in Santiago de Compostela for providing a stimulating environment where part of this work was carried out. TR thanks the Max-Planck-Institut f\"ur Physik for hospitality during the finalisation of this letter. In addition we would like to thank Ana L. Retore, Giacomo Piccinini and Sibylle Driezen for discussions. We also express our gratitude towards Alessandra Gnecchi, Daniel C. Thompson and  Dieter L\"ust for comments on the draft of this letter.

\bibliographystyle{JHEP}
\bibliography{Intbdefects_YB.bib}

\end{document}